\documentclass[a4paper]{jpconf}
\usepackage{graphicx}
\usepackage{multirow}
\begin{document}

\title{The $S_{3}$ symmetry: Flavour and texture zeroes}

\author{F. Gonz\'alez Canales$^{*}$ and A. Mondrag\'on$^{**}$}

\address{Instituto de F\'{\i}sica, Universidad Nacional Aut\'onoma de M\'exico, 04510, 
  M\'exico D.F., M\'exico}

\ead{$^{*}$ffelix@fisica.unam.mx, $^{**}$mondra@fisica.unam.mx}

\begin{abstract}
 We use the permutational symmetry group $S_{3}$ as a symmetry of flavour,  which leads to a 
 unified  treatment of masses  and mixings of the quarks and leptons. In this framework all mass 
 matrices of  the fermions in the theory  have the same form with four texture zeroes of class 
 I.  Also, with  the  help of six elements of real matrix representation of $S_ {3}$ as 
 transformation matrices of similarity  classes, we make a classification of the sets of 
 mass matrices with texture zeroes in equivalence classes. This classification  reduce the 
 number of  phenomenologically viable textures for the  non-singulars mass matrices of 
 $3\times3$, from thirty three down to only eleven independent sets of  matrices . 
 Each of these sets of  matrices has exactly the same physical content.
\end{abstract}

\section{Introduction}
The Standard Model (SM) can be extended by adding of three right-handed neutrino states,  which 
would be singlets of the gauge group of SM, but coupled to matter just through the neutrino 
masses~\cite{Xing:2010zz}. But in minimal extensions of the Standard Model, considering a mass 
term for left-handed neutrinos purely  of Dirac  is not theoretically favored,  because it can 
not explain easily why neutrinos are much lighter than the charged leptons. Thus, we assume that 
neutrinos have Majorana masses and acquire their small masses through of the type I seesaw 
mechanism. 

In both lepton and quark sectors of the extended Standard Model, analogous fermions in different 
generations, say ${\it u,c}$ and ${\it t}$ or ${\it d,s}$ and ${\it b}$, have completely 
identical couplings to all gauge bosons of the strong, electroweak interactions. Prior to the 
introduction of the Higgs boson and mass terms, the Lagrangian is chiral and invariant with 
respect to any permutation of the left and right quark fields. The introduction of a Higgs boson 
and the Yukawa couplings  give mass to the quarks and leptons when the gauge symmetry is 
spontaneously broken. The  mass term in the Lagrangian, obtained by taking the vacuum 
expectation value of the Higgs field in the quark and lepton Higgs couplings, gives rise to mass 
matrices ${\bf M_d}$ , ${\bf  M_u}$, ${\bf  M_l}$ and ${\bf  M_{\nu} }$;
\begin{equation}\label{eq:21}
{\cal L}_{Y} ={\bf \bar{q}}_{d,L}{\bf M}_{d}{\bf q}_{d,R}
 + {\bf\bar{q}}_{u,L}{\bf M}_{u}{\bf q}_{u,R}
 + {\bf \bar{L}}_{L}{\bf M}_{l}{\bf L}_{R}
 + {\bf \bar{\nu}}_{L}{\bf M}_{\nu} \left( {\bf \nu}_{L} \right)^{c}
 +h.c.
\end{equation}
Therefore,  we propose, as well as many other authors~\cite{Fritzsch:1977za,Mondragon:1998gy,
Barranco:2010we,Fritzsch:1999ee,Kubo:2003iw}, that the texture zeroes of the mass matrices of 
the quarks and leptons, are the result of a flavour permutational symmetry $S_{3}$ and its 
spontaneous or explicit breaking. 

On the other hand, in the last ten years, important theoretical advances have been made in the 
understanding of the mechanisms for the mass fermion  generation and flavour mixing. 
Phenomenologically, some striking progress has been made with the help of the texture zeroes and 
flavour symmetries in specifying the quantitative relationship between flavour mixing angles and 
quark or lepton mass ratios~\cite{Mondragon:1998gy,Barranco:2010we,Xing:2003zd,Fritzsch:1999ee} 
with a minimum of free parameters. In fact it, can be noted that different mass matrices  with 
texture zeroes located in different positions may have  exactly the same physical  content. 
Therefore the question arises, is there any relation between these  matrices?.  We find an 
answer to this question through the similarity clases, recently proposed by 
Branco~\cite{Branco:2007nn}.\\

In this paper, we use the permutational  symmetry $S_{3}$ as a flavour symmetry, in a unified 
treatment of masses and mixings of quarks and leptons.  Also, with the help of a real 
matrix representation of the group $S_ {3}$, as a basis for the transformation matrices 
of the similarity classes, we make a classification of the set of mass matrices with texture  
zeroes.

\section{Flavour permutational symmetry $S_{3}$}
A phenomenologically and theoretically meaningful approach for reducing the number of free 
parameters in the Standard Model  is the imposition of  texture 
zeroes~\cite{Fritzsch:1999ee,Xing:2003zd} or flavour symmetries. Recent flavour symmetry models 
are reviewed in~\cite{Altarelli:2009gn}. Also, certain texture zeroes may be obtained from a 
flavour symmetry. In particular, a permutational $S_{3}$ flavour symmetry and its sequential 
explicit breaking justifies taking the same generic form for the mass matrices of all Dirac 
fermions, conventionally called a four texture zeroes 
form~\cite{Mondragon:1998gy,Barranco:2010we}. Some reasons to propose the validity of a matrix 
with four texture zeroes as a universal form for the mass matrices of all Dirac fermions in the 
theory are the following:
\begin{enumerate}
 \item The idea of $S_{3}$ flavour symmetry and its explicit breaking has  been succesfully  
  realized as a mass matrix with four texture zeroes in the quark sector to interpret the strong 
  mass hierarchy of up and down type quarks~\cite{Fritzsch:1977za, Mondragon:1998gy}. Also, the 
  numerical values of mixing matrices of the quarks determined in this framework are in good 
  agreement with the experimental data~\cite{Mondragon:1998gy}.
 \item Since the mass spectrum of the charged leptons exhibits a hierarchy similar to the
  quark's one, it would be natural to consider the same $S_{3}$ symmetry and its explicit
  breaking to justify the use of the same generic form with four texture zeroes for the 
  charged lepton mass matrix.
 \item As for the Dirac neutrinos, we have no direct information about the absolute values   or 
  the relative values of the Dirac neutrino masses, but the mass matrix with four texture   
  zeroes can be obtained from an grand unified theory $SO(10)$ which describes well  the data  
  on masses and mixings of Majorana neutrinos~\cite{Buchmuller:2001dc}.  Furthermore,  from 
  supersymmetry arguments, it would be sensible to assume that the Dirac neutrinos have a  mass 
  hierarchy similar to that of the u-quarks and it would be natural to take for the  Dirac 
  neutrino mass matrix also a matrix with four texture zeroes.
\end{enumerate}
\subsection{Mass matrices from the breaking of $S_{3L}\otimes S_{3R}$ }
Some authors have pointed out that realistic Dirac fermion mass matrices 
results from the flavour permutational symmetry  $S_{L}(3)\otimes S_{R}(3)$ and its spontaneous 
or explicit breaking~\cite{Fritzsch:1977za,Mondragon:1998gy,Barranco:2010we,Fritzsch:1999ee,
Kubo:2003iw}. The group $S_{3}$ treats three treats three objects symmetrically, while its 
$3\times3$ representation structure $\bf{1\oplus2}$ treats the generations differently and 
adapts itself readily to the hierarchical nature of the mass spectrum. Under exact 
$S_{3L} \otimes S_{3R}$ symmetry, the mass spectrum for either quark sector (up or down quarks) 
or leptonic sector (charged leptons or Dirac neutrinos) consists of one massive particle in a 
singlet irreducible representation and a pair of massless particles in a doublet irreducible 
representation, the corresponding quark mass matrix with the exact $S_{3L} \otimes S_{3R}$ 
symmetry will be denoted by ${\bf M}_{i3}$ with $i = u, d, l, \nu_{_D}$. Here, $u$, $d$, $l$ and 
$\nu_{_D}$ denote the up  quarks, down quarks, charged leptons  and Dirac neutrinos, 
respectively. Assuming that there is only one Higgs boson in the theory, this 
$SU(2)_{L}$ doublet can be accommodated in a singlet representation of $S_{3}$.In order to 
generate masses for the first and second families, we add the terms ${\bf M}_{i2}$ and 
${\bf M}_{i1}$ to ${\bf M}_{i3}$. The term ${\bf M}_{i2}$ breaks the permutational symmetry  
$S_{3L} \otimes S_{3R}$ down to $S_{2L} \otimes S_{2R}$ and mixes the singlet and doublet 
representation of $S_{3}$. ${\bf M}_{i1}$ transforms as the mixed symmetry term in the doublet 
complex tensorial representation of $S^{diag}_{3} \subset S_{3L} \otimes S_{3R}$. Putting the 
first family in a complex representation will allow us to have a CP violating phase in the 
mixing matrix. Then, in a symmetry adapted basis, ${\bf M}_{i}$ takes the form
\begin{displaymath}
 \begin{array}{l}
 {M_{i}}={m_{i3}} 
  \left[ \pmatrix{
   0 & |{A_{i}}|e^{i\phi_{i}} & 0 \cr
   |{A_{i}}|e^{-i\phi_{i}} & 0 & 0 \cr
   0 & 0 & 0 \cr } 
  + \pmatrix{
   0 & 0 & 0 \cr
   0 & -\triangle_{i}+\delta_{i} & C_{i} \cr
   0 & C_{i} & \triangle_{i}-\delta_{i} \cr }
  + \pmatrix{
   0 & 0 & 0 \cr
   0 & 0 & 0 \cr
   0 & 0 & 1 - \triangle_{i} \cr } \right] 
 \end{array}
\end{displaymath}
\begin{equation}\label{eq:2.11}
\begin{array}{c}
  {M_{i}} = m_{i3} 
   \pmatrix{
    0 & A_{i} & 0 \cr
    A_{i}^{*} & B_{i} & C_{i} \cr
    0 & C_{i} & D_{i}  \cr }.
\end{array}
\end{equation}
where $A_{i} = |A_{i}|e^{i\phi_{i}}$, $B_{i} = -\triangle_{i} + \delta_{i}$ and 
$D_{i} = 1 - \delta_{i}$. From the strong hierarchy in the masses of the Dirac fermions, 
$m_{i3} >> m_{i2} > m_{i1}$, we expect $1-\delta_{i}$ to be very close to unity. The small 
parameter $\delta_{i}$ is a function of the flavour symmetry breaking parameter 
${Z_{i}}^{1/2}$~\cite{Mondragon:1998gy,Barranco:2010we}. In other words, each possible  symmetry 
breaking  pattern is now characterized by the flavour symmetry breaking parameter 
${Z_{i}}^{1/2}$, which is defined as the ratio 
{\small ${Z_{i}}^{1/2} = \frac{ \left( {M_{i}} \right)_{23} }{ \left({M_{i}} \right)_{22} }$}. 
This ratio measures the mixing of singlet and doublet irreducible representations of $S_{3}$. 
The mass matrix (\ref{eq:2.11}) may be written as 
$M_{ i } = P^{\dagger}_{ i } \bar{M}_{ i } P_{ i }$, where $\bar{M}_{ i }$ is a real symmetric 
matrix  and $P_{ i }\equiv \textrm{diag}\left[ 1, e^{ i\phi_{ i } }, e^{ i\phi_{ i } } \right]$.
\section{Classification  of texture zeroes in equivalence classes}
 In this section we make a classification of mass matrices with texture  zeroes in terms of  the 
 similarity clases. The similarity classes are defined as follows: Two matrices $M$ and $M'$ are 
 similar if there exists an  invertible matrix $T$ such that
\begin{equation}\label{similar:trans}
 M' = T M T^{-1} \quad \textrm{or} \quad M' = T^{-1} M T.
\end{equation}
The equivalence classes associated with a similarity transformation are called similarity 
classes. Another way to see the similarity classes is that the matrices that satisfy the 
similarity transformation, eq.~(\ref{similar:trans}), have the same invariants; Trace, 
determinant and $\chi$:
\begin{equation}\label{invariantes:1}
 \begin{array}{l}
  \textrm{Tr} \left \{ M \right \} = \textrm{Tr} \left \{ M' \right \}, \quad 
  \textrm{det} \left \{ M \right \} = \textrm{det} \left \{ M' \right \} \quad 
  \textrm{and} \quad 
  \chi' = \chi \equiv 
  \frac{ \textrm{Tr} \left \{ M^{2} \right \} - \textrm{Tr} \left \{ M \right \}^{2} }{2}.
 \end{array}
\end{equation} \\
In this paper we will count  the texture zeroes in a matrix as follows:  two texture zeroes off-
diagonal counts as one zero, while one on the diagonal counts as one 
zero~\cite{Fritzsch:1999ee}. But in the literature we find that a mass matrix has double number 
of texture zeroes than the number that we obtain with our rule. This is so, because in the 
literature the total number of texture zeroes counted in a mass matrix is the sum of the texture 
zeroes coming from the two mass matrices in the sector of  quarks  (up and down quarks) or 
leptons (charged leptons and left-handed neutrinos). Hence, to avoid confusion in the 
nomenclature  of the matrices, we count the number of texture zeroes in a matrix  with the rule 
previously enunciated, but when referring to this matrix we will follow the literature's rule   
and name it with the double of texture zeroes it actually has.\\
Now, from the  most general form of the mass matrices of $3\times3$, symmetric and  Hermitian:
 \begin{equation}\label{matrices:Herm:Sime}
  M^{ ^{\textrm{s}} } =
  \left( \begin{array}{ccc}
   g & a & e \\
   a & b & c \\
   e & c & d  
  \end{array} \right)
  \quad \textrm{and} \quad
   M^{ ^{\textrm{h} } } =
  \left( \begin{array}{ccc}
   g & a & e \\
   a^{*} & b & c \\
   e^{*} & c^{*} & d  
  \end{array} \right),
 \end{equation}
we can see that only six of the nine elements of these matrices  are independent of each other. 
Therefore, in a certain sense the similarity transformation, eq.~(\ref{similar:trans}), realized 
the permutation  of the six independent elements  in the nine entries of the mass matrices. But 
if we want to preserve the invariants~(\ref{invariantes:1}),  the elements on the  diagonal can 
only exchanged positions on the diagonal, while the  off-diagonal elements can only exchange 
positions outside of the diagonal. Thus we have that all these operations reduce to the 
permutations of three objects.  So it is natural propose to as transformation matrices $T$  in 
the similarity clases, see~eq.(\ref{similar:trans}), the six elements of the real representation 
of the group of  permutations $S_ {3}$, which are:
\begin{equation}\label{Base:completa}
  \begin{array}{l}
T\left( A_0 \right)=
\left( \begin{array}{ccc}
1& 0 & 0 \\
0 & 1 & 0 \\
0 & 0 & 1
\end{array}\right), 
T\left(A_1\right)=
\left( \begin{array}{ccc}
0 & 1 & 0 \\
1 & 0 & 0 \\
0 & 0 & 1
\end{array}\right), 
T\left(A_2\right)=
\left( \begin{array}{ccc}
0 & 0 & 1 \\
0 & 1 & 0 \\
1 & 0 & 0
\end{array}\right),\\
T\left(A_3\right)=
\left(  \begin{array}{ccc}
1 & 0 & 0 \\
0 & 0 & 1 \\
0 & 1 & 0
\end{array}\right), 
T\left(A_4\right)=
\left(  \begin{array}{ccc}
0 & 1 & 0 \\
0 & 0 & 1 \\
1 & 0 & 0
\end{array}\right), 
T\left(A_5\right)=
\left(  \begin{array}{ccc}
0 & 0 & 1 \\
1 & 0 & 0 \\
0 & 1 & 0
\end{array}\right).
\end{array} 
\end{equation} 
Then, applying  the transformations~(\ref{similar:trans}) and taking the 
matrices~(\ref{Base:completa}) as matrices of transformation, we get the classification of mass 
matrices with texture zeroes, which is shown in the tables~1,~2,~3 and 4. In this tables, the 
"${\star}$" and "$\times$" denote an arbitrary non-vanishing matrix element on the diagonal 
and off-diagonal entries, respectively.  
\begin{table}[h]
{\footnotesize \begin{center}
\begin{tabular}{|c|c|l|l|}
\hline
\multirow{2}{*}{Class }& \multirow{2}{*}{Textures} & \multicolumn{2}{|c|}{Invariants} \\ \cline{3-4}
& & Symmetric & Hermitian \cr \hline
I &  $
 \left( \begin{array}{ccc}
  0 & \times & 0 \\
  \times & 0 & 0 \\
  0 & 0 & \star  
 \end{array}  \right)
 \left( \begin{array}{ccc}
  \star & 0 & 0 \\
   0 & 0 & \times \\
   0 & \times & 0 
 \end{array}  \right)
 \left( \begin{array}{ccc}
   0 & 0 & \times \\
   0 & \star & 0 \\
   \times & 0 & 0 
 \end{array}  \right)$  & 
 $\begin{array}{l} 
   \textrm{Tr}   = d \\
   \textrm{det}  = -a^2d   \\
   {\bf \chi } = a^2
  \end{array}
  $ & $\begin{array}{l} 
   \textrm{Tr} = d \\
   \textrm{det} = -|a|^2d   \\
   {\bf \chi } = |a|^2
  \end{array} $ \\ \hline
\end{tabular}
\end{center} }
 \caption{\label{tabla1} Matrix with eight texture zeroes.} 
\end{table} 
\begin{table}[h]
{\footnotesize \begin{center}
\begin{tabular}{|c|c|l|l|}
\hline
\multirow{2}{*}{Class }& \multirow{2}{*}{Textures} & \multicolumn{2}{|c|}{Invariants} \\ \cline{3-4}
& & Symmetric & Hermitian \cr \hline
I &  $ \begin{array}{l} 
 \left( \begin{array}{ccc}
  \star & 0 & 0 \\
  0 & 0 & \times \\
  0 & \times & \star  
 \end{array}  \right)
 \left( \begin{array}{ccc}
   0 & 0 & \times \\
   0 & \star & 0 \\
   \times & 0 & \star 
 \end{array}  \right)
 \left( \begin{array}{ccc}
   \star & \times & 0 \\
   \times & 0 & 0 \\
    0 & 0 & \star  
 \end{array}  \right)
  \\
  \left( \begin{array}{ccc}
  \star & 0 & 0 \\
   0 & \star & \times \\
   0 & \times & 0 
 \end{array}  \right)
 \left( \begin{array}{ccc}
  0 & \times & 0 \\
  \times & \star & 0 \\
   0 & 0 & \star 
 \end{array}  \right)
 \left( \begin{array}{ccc}
   \star & 0 & \times \\
   0 & \star & 0 \\
   \times & 0 & 0 
 \end{array}  \right)
\end{array} $  
  & 
 $\begin{array}{l} 
   \textrm{Tr}   = d + g \\
   \textrm{det}  = -c^2g   \\
   {\bf \chi } = c^2 -gd
  \end{array}
  $ & $\begin{array}{l} 
   \textrm{Tr} = d + g \\
   \textrm{det} = -|c|^2g   \\
   {\bf \chi } = |c|^2 - gd
  \end{array} $ \\ \hline
  II &  $ \begin{array}{l} 
 \left( \begin{array}{ccc}
  0 & \times & 0 \\
  \times & 0 & \times \\
  0 & \times & \star 
 \end{array}  \right)
 \left( \begin{array}{ccc}
   0 & \times & \times \\
   \times & 0 & 0 \\
   \times & 0 & \star 
 \end{array}  \right)
 \left( \begin{array}{ccc}
   \star & \times & 0 \\
   \times & 0 & \times \\
    0 & \times & 0  
 \end{array}  \right)
  \\
  \left( \begin{array}{ccc}
   0 & 0 & \times \\
   0 & \star & \times \\
   \times & \times & 0 
 \end{array}  \right)
 \left( \begin{array}{ccc}
  0 & \times & \times \\
  \times & \star & 0 \\
   \times & 0 & 0 
 \end{array}  \right)
 \left( \begin{array}{ccc}
   \star & 0 & \times \\
   0 & 0 & \times \\
   \times & \times & 0 
 \end{array}  \right)
\end{array} $ 
  & 
 $\begin{array}{l} 
   \textrm{Tr}   = d  \\
   \textrm{det}  = -a^2d   \\
   {\bf \chi } = a^2 + c^2 
  \end{array}
  $ & $\begin{array}{l} 
   \textrm{Tr} = d \\
   \textrm{det} = -|a|^2d   \\
   {\bf \chi } = |a|^2 +|c|^2 
  \end{array} $ \\ \hline
  III &  $ \begin{array}{l} 
 \left( \begin{array}{ccc}
  \star & 0 & 0 \\
  0 & \star & 0 \\
  0 & 0 & \star 
 \end{array}  \right)
 \end{array} $  
  & 
 $\begin{array}{l} 
   \textrm{Tr}   = g + b + d  \\
   \textrm{det}  = g  b  d   \\
   {\bf \chi } = -gb - gd \\ \quad -bd
  \end{array}
  $ & $\begin{array}{l} 
   \textrm{Tr}   = g + b + d  \\
   \textrm{det}  = g  b  d   \\
   {\bf \chi } = -gb - gd \\ \quad -bd
  \end{array} $ \\ \hline
  IV &  $ \begin{array}{l} 
 \left( \begin{array}{ccc}
  0 & \times & \times \\
  \times & 0 & \times \\
  \times & \times & 0 
 \end{array}  \right)
 \end{array} $ 
  & 
 $\begin{array}{l} 
   \textrm{Tr}   = 0 \\
   \textrm{det}  = 2ace   \\
   {\bf \chi } = a^2 + e^2 \\ \quad + c^2
  \end{array}
  $ & $\begin{array}{l} 
   \textrm{Tr}   = 0 \\
   \textrm{det}  = a^*c^*e \\ \quad + ace^*   \\
   {\bf \chi } = |a|^2 + |e|^2 \\ \quad + |c|^2
  \end{array} $ \\ \hline
\end{tabular}
\end{center}}
 \caption{\label{Tabla2} Matrix with six texture zeroes.} 
\end{table} 
\begin{table}[h]
{\footnotesize \begin{center}
\begin{tabular}{|c|c|l|l|}
\hline
\multirow{2}{*}{Class }& \multirow{2}{*}{Textures} & \multicolumn{2}{|c|}{Invariants} \\ \cline{3-4}
& & Symmetric & Hermitian \cr \hline
I &  $ \begin{array}{l} 
 \left( \begin{array}{ccc}
  0 & \times & 0 \\
  \times & \star & \times \\
  0 & \times & \star 
 \end{array}  \right)
 \left( \begin{array}{ccc}
   0 & 0 & \times \\
   0 & \star & \times \\
   \times & \times & \star  
 \end{array}  \right)
 \left( \begin{array}{ccc}
   \star & 0 & \times \\
    0 & 0 & \times \\
    \times & \times & \star   
 \end{array}  \right)
  \\
  \left( \begin{array}{ccc}
  \star & \times & \times \\
   \times & 0 & 0 \\
   \times & 0 & \star 
 \end{array}  \right)
 \left( \begin{array}{ccc}
  \star & \times & 0 \\
  \times & \star & \times \\
   0 & \times & 0 
 \end{array}  \right)
 \left( \begin{array}{ccc}
   \star & \times & \times \\
   \times & \star & 0 \\
   \times & 0 & 0 
 \end{array}  \right)
\end{array} $  
  & 
 $\begin{array}{l} 
   \textrm{Tr}   = b + d  \\
   \textrm{det}  = -a^2d   \\
   {\bf \chi } = a^2 + c^2 \\ \quad -bd
  \end{array}
  $ & $\begin{array}{l} 
   \textrm{Tr}   = b + d  \\
   \textrm{det}  = -|a|^2d   \\
   {\bf \chi } = |a|^2 + |c|^2 \\ \quad -bd
  \end{array} $ \\ \hline
  II &  $ \begin{array}{l} 
 \left( \begin{array}{ccc}
  0 & \times & \times \\
  \times & \star & 0 \\
  \times & 0 & \star 
 \end{array}  \right)
 \left( \begin{array}{ccc}
   \star & \times & 0 \\
   \times & 0 & \times \\
    0 & \times & \star 
 \end{array}  \right)
 \left( \begin{array}{ccc}
   \star & 0 & \times \\
    0 & \star & \times \\
    \times & \times & 0  
  \end{array} \right)
  \end{array} 
 $ 
  & 
 $\begin{array}{l} 
   \textrm{Tr}   = b + d  \\
   \textrm{det}  = -a^2d \\ \quad - e^2b  \\
   {\bf \chi } = a^2 + e^2 \\ \quad -bd
  \end{array}
  $ & $\begin{array}{l} 
   \textrm{Tr}   = b + d  \\
   \textrm{det}  = -|a|^2d \\ \quad - |e|^2b  \\
   {\bf \chi } = |a|^2 + |e|^2 \\ \quad -bd
  \end{array} $ \\ \hline
  III & $ \begin{array}{l} 
 \left( \begin{array}{ccc}
  0 & \times & \times \\
  \times & 0 & \times \\
  \times & \times & \star 
 \end{array}  \right)
 \left( \begin{array}{ccc}
   0 & \times & \times \\
   \times & \star & \times \\
   \times & \times & 0 
 \end{array}  \right)
 \left( \begin{array}{ccc}
   \star & \times & \times \\
   \times & 0 & \times \\
   \times & \times & 0  
  \end{array} \right)
  \end{array} 
 $ 
  & 
 $\begin{array}{l} 
   \textrm{Tr}   = d \\
   \textrm{det}  = 2 ace - a^2d   \\
   {\bf \chi } = a^2 + c^2 \\ \quad + e^2
  \end{array}
  $ & $\begin{array}{l} 
  \textrm{Tr}   = d \\
   \textrm{det}  = a^*c^*e \\ \quad + ace^* - |a|^2d   \\
   {\bf \chi } = |a|^2 + |c|^2 \\ \quad + |e|^2
  \end{array} $ \\ \hline
  IV & $ \begin{array}{l} 
 \left( \begin{array}{ccc}
  \star & 0 & 0 \\
   0 & \star & \times \\
   0 & \times & \star 
 \end{array}  \right)
 \left( \begin{array}{ccc}
   \star & 0 & \times \\
    0 & \star & 0 \\
   \times & 0 & \star 
 \end{array}  \right)
 \left( \begin{array}{ccc}
   \star & \times & 0 \\
   \times & \star & 0 \\
    0 & 0 & \star   
  \end{array} \right)
  \end{array} 
 $ 
  & 
 $\begin{array}{l} 
   \textrm{Tr}   = g + b + d  \\
   \textrm{det}  = -gc^2 + gbd  \\
   {\bf \chi } = c^2 - gb \\ \quad -gd -bd
  \end{array}
  $ & $\begin{array}{l} 
   \textrm{Tr}   = g + b + d  \\
   \textrm{det}  = -g|c|^2 - gbd  \\
   {\bf \chi } = |c|^2 - gb \\ \quad -gd -bd
  \end{array} $ \\ \hline
\end{tabular}
\end{center}}
 \caption{\label{tabla3} Matrix with four texture zeroes.} 
\end{table} 

\begin{table}[h]
{ \footnotesize \begin{center}
\begin{tabular}{|c|c|l|l|}
\hline
\multirow{2}{*}{Class }& \multirow{2}{*}{Textures} & \multicolumn{2}{|c|}{Invariants} \\ \cline{3-4}
& & Symmetric & Hermitian \cr \hline
I &  $\begin{array}{c}
 \left( \begin{array}{ccc}
  0 & \times & \times \\
  \times & \star & \times \\
  \times & \times & \star 
 \end{array}  \right)
 \left( \begin{array}{ccc}
  \star & \times & \times \\
  \times & 0 & \times \\
  \times & \times & \star  
 \end{array}  \right)
 \left( \begin{array}{ccc}
   \star & \times & \times \\
   \times & \star & \times \\
   \times & \times & 0 
 \end{array}  \right) \\
 \end{array} $  & 
 $\begin{array}{l} 
   \textrm{Tr}   = b + d \\
   \textrm{det}  = 2ace -a^2d \\ \quad-be^2   \\
   {\bf \chi } = a^2 + c^2 \\ \quad + e^2 -bd
  \end{array}
  $ & $\begin{array}{l} 
    \textrm{Tr}   = b + d \\
   \textrm{det}  =  -|a|^2d -b|e|^2 \\  +ace^* + a^*c^*e \\
   {\bf \chi } = |a|^2 + |c|^2 + |e|^2 \\ \quad -bd
  \end{array} $ \\ \hline
  II &  $
\begin{array}{c} 
 \left( \begin{array}{ccc}
  \star & 0 & \times \\
   0 & \star & \times \\
  \times & \times & \star 
 \end{array}  \right)
 \left( \begin{array}{ccc}
  \star & \times & \times \\
  \times & \star & 0 \\
  \times &  0 & \star  
 \end{array}  \right)
 \left( \begin{array}{ccc}
   \star & \times & 0 \\
   \times & \star & \times \\
   0 & \times & \star 
 \end{array}  \right)
 \end{array}
  $  & 
 $\begin{array}{l} 
   \textrm{Tr}   = b + d + g \\
   \textrm{det}  = gbd \\ \quad - c^2g - e^2 b  \\
   {\bf \chi } = e^2 + c^2 -gb\\ \quad  -gd -bd
  \end{array}
  $ & $\begin{array}{l} 
   \textrm{Tr}   = b + d + g \\
   \textrm{det}  = gbd  \\ \quad- |c|^2g - |e|^2 b  \\
   {\bf \chi } = |e|^2 + |c|^2 -gb \\ \quad -gd -bd
  \end{array} $ \\ \hline
\end{tabular} 
\end{center} }
 \caption{\label{tabla4} Matrix with two texture zeroes.} 
\end{table} 
\section{Seesaw mechanism}
The left-handed Majorana neutrinos acquire their small masses through the type I seesaw 
mechanism $M_{ \nu_{L} } = M_{ \nu_{D} } M_{ \nu_R }^{-1} M_{ \nu_{D} }^{T}$, where 
$M_{ \nu_{D} }$ and $ M_{ \nu_{R} }$ denote the Dirac and right handed Majorana neutrino mass 
matrices, respectively. The form of $M_{ \nu_{D} }$ is given in eq.~(\ref{eq:2.11}); which is a 
matrix with four texture zeroes class  I, Hermitian, and from our conjecture of a universal 
$S_{3}$ flavour symmetry in a unified treatment of all fermions, it is natural to take for 
$M_{ \nu_R }$ also a matrix with four texture zeroes of class I, but symmetric. Let us further 
assume that the phases in the entries of the  $M_{ \nu_{ R} }$  may be factorized out as 
$ M_{ \nu_{ _R } } = R \bar{M}_{ \nu_{ _R } }R$, where 
$R \equiv \textrm{diag}\left[ e^{ - i\phi_{ c }}, e^{ i\phi_{ c } }, 1 \right]$ with 
$\phi_{ c } \equiv \arg \left \{ c_{ \nu_{ _R } }  \right \}$ and
\begin{equation}
 \bar{M}_{ \nu_{ _R } } = \left( \begin{array}{ccc} 
   0 &  a_{ \nu_{ _R } }  & 0  \\
   a_{ \nu_{ _R } }   & | b_{ \nu_{ _R } } |    &| c_{ \nu_{ _R } } | \\
   0 &  | c_{ \nu_{ _R } }  | & d_{ \nu_{ _R } }   
 \end{array}\right),
\end{equation}
Then, the type I seesaw mechanism takes the form $M_{  \nu_{ _L }  } = P_{ _D }^{\dagger}  
\bar{M}_{\nu_{ _D } } P_{ _D } R^{ \dagger } \bar{M}_{ \nu_{ _R } } ^{-1} R^{ \dagger } 
P_{ _D } \bar{M}_{\nu_{ _D } } P_{ _D }^{ \dagger }$ and the mass matrix of the left-handed Majorana 
neutrinos has the following form, with four texture zeroes of class I: 
\begin{equation}\label{seesaw:F}
  M_{ \nu_{ _L } } =  \left( \begin{array}{ccc} 
   0 &  a_{ \nu_{ _L } } & 0  \\
   a_{ \nu_{ _L } } & b_{ \nu_{ _L } } & c_{ \nu_{ _L } }  \\
   0 & c_{ \nu_{ _L } } & d_{ \nu_{ _L } }   
  \end{array}\right),
\end{equation}
where
{\small \begin{equation}\label{seesaw:F:elem}
 \begin{array}{l}
  a_{ \nu_{ _L } } = \frac{ | a_{ \nu_{ _D } } |^{2} }{  a_{ \nu_{ _R } } } ,
  \quad   d_{ \nu_{ _L } } = \frac{ d_{ \nu_{ _D } }^{2} }{ d_{ \nu_{ _R } } },
  \quad c_{ \nu_{ _L } } = \frac{ c_{ \nu_{ _D } } d_{ \nu_{ _D } } }{ d _{ \nu_{ _R } } }+  
   \frac{ | a_{ \nu_{ _D } } | }{   | a_{ \nu_{ _R } } |}
  \left ( c_{ \nu_{ _D } }e^{- i \phi_{ \nu_{ D } } } - \frac{ |c_{ \nu_{ _R } }| d_{ \nu_{ 
  _D } }  }{ d_{ \nu_{ _R } 
  } } e^{ i \left( \phi_{ c } - \phi_{ \nu_{ D } }  \right) } \right) 
  \\ \\
  b_{ \nu_{ _L } } = \frac{ c_{ \nu_{ _D } }^{ 2 } }{ d_{ \nu_{ _R } } }  + \frac{ | c_{ 
  \nu_{ _R } }|^{2} -  | b_{ \nu_{ _R } }|  d_{ \nu_{ _R } } }{  d_{ \nu_{ _R } } } 
  \frac{ | a_{ \nu_{ _D } } |^{ 2 } }{  a_{ \nu_{ _R } }^{2} } e^{ i 2\left( \phi_{ c } -
   \phi_{ \nu_{ D } } \right)}   
    + 2 \frac{ | a_{ \nu_{ _D } }| }{ |a_{ \nu_{ _R } } | } \left( b_{ \nu_{ _D } } 
  e^{- i \phi_{ \nu_{ D } } } 
  - \frac{c_{ \nu_{ _D } } |c_{ \nu_{ _R } }| }{ d_{ \nu_{ _R } }  }  e^{ i \left( \phi_{ c 
  } - \phi_{ \nu_{ D } }   \right) } \right)   . 
  \end{array}
\end{equation} }
The elements $a_{ \nu_{ _L } }$ and $d_{ \nu_{ _L } }$ are real , while $b_{ \nu_{ _L } }$ and 
$c_{ \nu_{ _L } }$ are complex.  It may also be noticed that $M_{ \nu_{ _L } }$ may also have four 
texture zeroes of class I when $M_{ \nu_{ _R } }$ has four texture zeroes of class I, six texture 
zeroes of class II, six texture zeroes of class I and eight texture zeroes of class I. From ecs.~(\ref{seesaw:F:elem}) 
we conclude that the information of the number of texture zeroes in $M_{ \nu_{ _R } }$ is found in 
elements (2,2) and (2,3) of the matrix $M_{ \nu_{ _L } }$. In this case, without loss of generality, 
2we may choose 
$\arg \left\{  b_{ \nu_{ _L } } \right \} = 2\arg \left\{ c_{ \nu_{ _L } } \right \} = 2\varphi$, 
the analysis simplifies since the phases in $M_{ \nu_{ _L } }$ may be factorized as 
$ M_{ \nu_{ _L } } = Q \bar{M}_{ \nu_{ _L } } Q$, where $Q$ is a diagonal  matrix of phases
{\small $Q \equiv \textrm{diag}\left[ e^{ -i  \varphi }, e^{ i \varphi }, 1 \right] $ }
and $\bar{M}_{ \nu_{ _L } }$ is a real symetric matrix. 
\section{Mass matrix with four texture zeroes as function of the fermion masses }
Now, computing the invariants (\ref{invariantes:1}) of the real symetric matrix 
$\bar{M}_{ \nu_{ _L } }$, we may express the parameters $A_{ i }$, $B_{ i }$, $C_{ i }$ and 
$D_{ i }$ 
occuring in (\ref{eq:2.11}) in terms of the 
mass eigenvalues. In this  way,  we get the $\bar{M}_{ i }$ matrix $(i=u,d,l,\nu_{ _L })$,  
reparametrized in terms of its eigenvalues and the parameter $\delta_{ i } $ is
\begin{equation}\label{T_fritzsch:ev}
 \bar{ M  }_{ i } =  \left( \begin{array}{ccc} 
   0 & \sqrt{ \frac{ \widetilde{ m }_{ i1 }   
    \widetilde{ m }_{ i2 } }{ 1 -  \delta_{ i } } } & 0 \\ 
   \sqrt{ \frac{ \widetilde{ m }_{ i1 }   \widetilde{ m }_{ i2 } }{  1 - \delta_{ i } } } & 
   \widetilde{ m }_{ i1 }  - 
   \widetilde{ m }_{ i2 } +  \delta_{ i } & 
   \sqrt{ \frac{\delta_{ i } }{ ( 1 - \delta_{ i } ) } f_{ i1 }  f_{ i2 } }   \\ 
   0 & \sqrt{ \frac{ \delta_{ i } }{ ( 1 - \delta_{ i } ) } f_{ i1 } f_{ i2 } } &  
   1 - \delta_{ i }
 \end{array}\right),
\end{equation}
where  $\widetilde{m}_{i1} = \frac{ m_{i1} }{ m_{i3} }$,  
$\widetilde{m}_{i2} = \frac{ | m_{i2} | }{ m_{i3} }$, 
$f_{i1}=1-\widetilde{m}_{i1}-\delta_{i}$ and $f_{i2} =1+ \widetilde{m}_{i2}  - \delta_{i}$. 
The small parameters $\delta_{i}$ are also functions of the mass ratios and the flavour
symmetry breaking parameter $Z^{1/2}_{i}$. The small parameter $\delta_{i}$ is obtained as the solution 
of the cubic equation {\small $( 1 - \delta_{ i } ) (\widetilde{ m }_{ i1 } - \widetilde{ m }_{ i2 } +  
\delta_{ i } )^{2} Z - \delta_{ i } f_{ i1 }  f_{ i2 } =0,$} and may be written as
\begin{equation}
 \delta_{i} = \frac{ Z_{ i } }{ Z_{ i } + 1 } 
 \frac{ \left( \widetilde{m}_{i2} - \widetilde{m}_{i1} \right)^{2} }{ W _{i}\left( Z \right) }
\end{equation}
where $W _{i}\left( Z \right)$ is the product of the two roots that do not vanish when $Z_{i}$ 
vanishes
{\footnotesize \begin{displaymath}
 \begin{array}{l}
  W _{i}\left( Z \right)  = \left[ p^{3}_{i} + 2 q^{2}_{i} + 2q \sqrt{ p^{3}_{i} +
   q^{2}_{i} } \right]^{ \frac{ 1 }{ 3 } } - | p_{i} |   
  + \left[ p^{3}_{i} + 2 q^{2}_{i} - 2q_{i} \sqrt{ p^{3}_{i} + q^{2}_{i} } 
  \right]^{ \frac{ 1 }{ 3 } } + \\
  + \frac{1}{9} \left( Z_{i} \left( 2 \left( \widetilde{m}_{i2} - 
  \widetilde{m}_{i1}\right) + 1 \right) + \left( \widetilde{m}_{i2} - 
  \widetilde{m}_{i1}\right) + 2 \right)^{2} 
  -\frac{1}{3} \left( \left[ q_{i} + \sqrt{ p^{3}_{i} + q^{2}_{i} } 
  \right]^{ \frac{ 1 }{ 3 } } + 
  \left[ q_{i}- \sqrt{ p^{3}_{i} + q^{2}_{i} } \right]^{ \frac{ 1 }{ 3 } } \right) 
  \times \\ \times 
  \left( Z_{i} \left( 2 \left( \widetilde{m}_{i2} - 
  \widetilde{m}_{i1}\right) + 1 \right) + \left( \widetilde{m}_{i2} - 
  \widetilde{m}_{i1}\right) + 2 \right)
 \end{array}
\end{displaymath} }
with {\small $p_{i} = -\frac{1}{3} \frac{ Z_{ i } \left( Z_{i} 
  \left( 2 \left( \widetilde{m}_{i2} - 
  \widetilde{m}_{i1}\right) + 1 \right) + \widetilde{m}_{i2} - \widetilde{m}_{i1} + 2 \right)^{2} 
   }{ Z_{ i } + 1 } + \frac{ \left[ Z_{i} \left( \widetilde{m}_{i2} - \widetilde{m}_{i1} \right) 
  \left( \widetilde{m}_{i2} - \widetilde{m}_{i1} + 2 \right) 
  \left(1 + \widetilde{m}_{i2} \right) \left( 1 - \widetilde{m}_{i1} \right) \right] }{ Z_{ i } + 1 } 
  ,$} and {\small $q_{i} = \frac{1}{6} \frac{ \left[ Z_{i} \left( \widetilde{m}_{i2} 
   - \widetilde{m}_{i1}  \right) \left( \widetilde{m}_{i2} - \widetilde{m}_{i1} + 2 \right) 
  \left(1 + \widetilde{m}_{i2} \right) \left( 1 - \widetilde{m}_{i1} \right) \right]
  \left( Z_{i} \left( 2 \left( \widetilde{m}_{i2} - \widetilde{m}_{i1} \right)+ 1 \right) 
  + \widetilde{m}_{i2} - \widetilde{m}_{i1} + 2 \right) }{ \left(  Z_{ i } + 1 \right)^{2} } 
  -\frac{1}{27} \frac{  \left( Z_{i} \left( 2   \left( \widetilde{m}_{i2} - \widetilde{m}_{i1}\right) + 1 \right) + \widetilde{m}_{i2} -\widetilde{m}_{i1} + 2 \right)^{3} }{ \left( Z_{ i } + 1 
  \right)^{ 3 } }. $} Also, the values allowed for the parameters $\delta_{i}$ are in the following 
  range $ 0 < \delta_{ i } < 1 - \widetilde{m}_{ i1 }$.  \\
Now, the entries in the real orthogonal  matrix ${\bf O}$ that diagonalize the matrix 
$\bar{M}_{ i }$, may also be expressed as
{\small \begin{equation}\label{M_ortogonal}
 {\bf O_{i}=}\left(\begin{array}{ccc}
  \left[ \frac{ \widetilde{m}_{i2} f_{i1} }{ {\cal D}_{ i1 } } \right]^{ \frac{1}{2} } & 
 -\left[ \frac{ \widetilde{m}_{i1} f_{i2} }{ {\cal D}_{ i2 } } \right]^{ \frac{1}{2} } & 
  \left[ \frac{ \widetilde{m}_{i1} \widetilde{m}_{i2} \delta_{i} }{ {\cal D}_{i3} }
   \right]^{ \frac{1}{2} } \\
  \left[ \frac{ \widetilde{m}_{i1} ( 1 - \delta_{i} ) f_{i1} }{ {\cal D}_{i1}  } 
   \right]^{ \frac{1}{2} } & 
  \left[ \frac{ \widetilde{m}_{i2} ( 1 - \delta_{i} ) f_{i2} }{ {\cal D}_{i2} } 
   \right]^{ \frac{1}{2} }  &  
  \left[ \frac{ ( 1 - \delta_{i} ) \delta_{i} }{ {\cal D}_{i3} } \right]^{ \frac{1}{2} } \\
 -\left[ \frac{ \widetilde{m}_{i1} f_{i2} \delta_{i} }{ {\cal D}_{i1} } 
   \right]^{ \frac{1}{2} } & 
 -\left[ \frac{ \widetilde{m}_{i2} f_{i1} \delta_{i} }{ {\cal D}_{i2} } 
   \right]^{ \frac{1}{2} } &
  \left[ \frac{ f_{i1} f_{i2} }{ {\cal D}_{i3} } \right]^{ \frac{1}{2} }
 \end{array}\right) ,
\end{equation} }
where, ${\cal D}_{i1} = ( 1 - \delta_{i} )( \widetilde{m}_{i1} + \widetilde{m}_{i2} ) 
  ( 1 - \widetilde{m}_{i1} )$, ${\cal D}_{i2} = ( 1 - \delta_{i} )( \widetilde{m}_{i1} 
  + \widetilde{m}_{i2} ) ( 1 + \widetilde{m}_{i2} ) $ and ${\cal D}_{i3} = ( 1 - \delta_{i} )
  ( 1 - \widetilde{m}_{i1} )( 1 + \widetilde{m}_{i2} )$.
\section{Mixing Matrices as Functions of the Fermion Masses}
 The quark mixing matrix $V_{_{CKM} }= U_{u}U_{d}^{\dagger}$ takes the form 
\begin{equation}\label{M_unitaria2}
 V_{_{CKM} }^{ ^{th} } =   {\bf O_{u} }^{T} P^{(u-d)}  {\bf O}_{d},
\end{equation}
where $P^{ (u-d) } = \textrm{diag}\left[1, e^{i\phi}, e^{i\phi} \right]$ with $\phi = \phi_{u} 
- \phi_{d}$, and $ {\bf O }_{ u,d }$ are the real orthogonal matrices~(\ref{M_ortogonal}) that 
diagonalize the real symmetric mass matrices $\bar{M}_{u,d}$. 
In a similar way, the lepton mixing matrix $U_{ _{PMNS} }= U_{l}^{\dagger}U_{\nu}$ may be written as 
\begin{equation}\label{M_unitaria3}
 U_{ _{PMNS } }^{ ^{th} } = { \bf O}_{l}^{T}P^{ ( \nu - l ) } {\bf O}_{\nu} K,
\end{equation}
where $P^{ ( \nu -l )} = \textrm{diag}\left[1, e^{ i \Phi_{1}  }, e^{ i \Phi_{2} }  \right]$ 
is the diagonal matrix of the Dirac phases, with $\Phi_{1} =  2\varphi - \phi_{ l }$ and 
$\Phi_{2} = \varphi - \phi_{l}$. The real orthogonal matrices ${ \bf O}_{ \nu, l } $ are 
defined in eq.~(\ref{M_ortogonal}). Exact explicit expressions for the unitary matices 
(\ref{M_unitaria2}) and (\ref{M_unitaria3}) are given by J. Barranco, F. Gonzalez Canales and A. 
Mondragon~\cite{Barranco:2010we}. 
\subsection{The  $\chi^{2}$ fit for the Quark Mixing Matrix}
We made a $\chi^{2}$ fit of the theoretical expressions for the modulii of the entries of the 
quark mixing matrix $| ( V_{ _{ CKM } }^{ ^{th} } )_{ij} |$,~eq.(\ref{M_unitaria2}) to the 
experimental values $| ( V_{ _{ CKM } }^{ ^{exp} } )_{ij} |$~\cite{Nakamura:2010zzi}. We 
computed the modulii of the entries of the quark mixing matrix and the inner angles of the 
unitarity triangle from the theoretical expresion~(\ref{M_unitaria2}) with the following 
numerical values of the quark mass ratios~\cite{Nakamura:2010zzi}:
\begin{equation}\label{quark-rat}
 \begin{array}{l}
  \widetilde{m}_{u} = 2.5469 \times 10^{ -5}, \; 
  \widetilde{m}_{c} = 3.9918 \times 10^{ -3}, \;
  \widetilde{m}_{d} = 1.5261 \times 10^{ -3}, \; 
  \widetilde{m}_{s} = 3.2319 \times 10^{ -2}.
 \end{array}
\end{equation} 
The resulting best values of the parameters $\delta_{u} = 3.829 \times 10^{-3}$, 
$\delta_{d} = 4.08 \times 10^{ -4 }$ and the Dirac CP violating phase is $\phi = 90^{o}$. 
The best values for the moduli of the entries of the $CKM$ mixing matrix are given in the 
following expresion
\begin{equation}
 \left| V_{ _{CKM} }^{ ^{th} } \right| =  \left(\begin{array}{ccc}
  0.97421 & 0.22560 & 0.003369 \\
  0.22545 & 0.97335 & 0.041736 \\
  0.008754 & 0.04094 & 0.99912  
 \end{array} \right)
\end{equation}
and the inner angles of the unitary triangle $\alpha^{ ^{th} } = 91.24^{o}$, 
$\beta^{ ^{th} } = 20.41^{o}$ and $\gamma^{ ^{th} } = 68.33^{o}$. The Jarlskog invariant takes 
the value $J_{q}^{ ^{th} } = 2.9 \times 10^{-5}$. All these results are in very good agreement 
with the experimental values.  The minimum value of $\chi^{2}$ obtained in this fit is 4.6 and 
the resulting value of $\chi^{2}$ for degree of freedom is 
{\small $\frac{\chi^{2}_{min} }{ d.o.f. }=0.77$}. In this way, we obtain the following numerical 
values for the mixing angles: $\theta_{12}^{q^{th}} = 13^{o}$, $\theta_{23}^{q^{th}} = 2.38^{o}$ 
and $\theta_{13}^{q^{th}} = 0.19^{o}$. Which are also in very good agreement with the latest 
analysis of the experimental data~\cite{Nakamura:2010zzi}. 
\subsection{ The $\chi^{2}$ fit for the Lepton Mixing Matrix}
In the case of the lepton mixing matrix, we made a $\chi^{2}$ fit of the theoretical 
expressions for the modulii of the entries of the lepton mixing matrix 
$| ( U_{ _{ PMNS } }^{ ^{th} } )_{ij} |$ to the  values extracted from experiment as given by 
Gonzalez-Garcia~\cite{GonzalezGarcia:2007ib}.The computation was made using the following values 
for the charged lepton masses~\cite{Nakamura:2010zzi}:
\begin{equation}\label{massChL}
\begin{array}{l}
 m_{e} = 0.5109~\textrm{MeV}, \;\; m_{\mu}= 105.685~\textrm{MeV}, \;\;   
 m_{\tau}=1776.99~\textrm{MeV}.
\end{array} 
\end{equation}
We took for the masses of the left-handed Majorana neutrinos a normal hierarchy. From the best 
values obtained for $m_{ \nu_{3} }$ and the experimental values of the  $\Delta m_{ 32 }^{ 2 }$ 
and $\Delta m_{ 21 }^{ 2 }$, we obtained the following best values for the neutrino masses
\begin{equation}\label{X2:Mnus}
 \begin{array}{l}
 m_{\nu_{1}} = 2.7  \times 10^{-3}\textrm{eV}, \quad  m_{\nu_{2}} =  9.1  \times 
 10^{-3}\textrm{eV}, \quad  m_{\nu_{3}} = 4.7 \times 10^{-2}\textrm{eV}. 
 \end{array}
\end{equation} 
The resulting best values of the parameters $\delta_{l} = 0.06$, $\delta_{ \nu } = 0.522$ and 
the Dirac CP violating phases are $\Phi_{1} = \pi \quad \textrm{and} \quad \Phi_{2} = 3\pi/2$.
The best values for the modulii of the entries of the $PMNS$ mixing matrix are given in the 
following expresion
\begin{equation}
 \left| U_{ _{PMNS} }^{ ^{th} } \right| =
 \left(\begin{array}{ccc}
  0.820421 & 0.568408 & 0.061817 \\
  0.385027 & 0.613436 & 0.689529 \\
  0.422689 & 0.548277 & 0.721615
 \end{array} \right).
\end{equation} 
The value of the rephasing invariant related to the Dirac phase is 
$J_{ l }^{ ^{th} } = 8.8 \times 10^{ -3}$. In this numerical analysis, the minimum value of the 
$\chi^{2}$, corresponding to the best fit, is $\chi^{2}=0.288$ and the resulting value of 
$\chi^{2}$ degree of freedom is {\small $\frac{\chi^{2}_{min} }{ d.o.f. }=0.075$}. All numerical 
results of the fit are in very good agreement with the values of the moduli of the entries in 
the matrix $U_{ _{PMNS} }^{ ^{exp} }$ as given in Gonzalez-Garcia~\cite{GonzalezGarcia:2007ib}. 
We obtain the following numerical values for the mixing angles:
 $\theta_{12}^{l^{th}} = 34.7^{o}$, $ \theta_{23}^{l^{th}} = 43.6^{o}$ and 
 $\theta_{13}^{l^{th}} =  3.5^{o}$. Which are also in very good agreement with the latest 
 experimental data~\cite{GonzalezGarcia:2007ib}.

\section{Conclusions}
In this work we have shown that, starting from the flavour permutational symmetry $S_{3}$, a 
simple and explicit ansatz about the pattern of symmetry breaking leads to a unified treatment  
of masses of quarks and leptons, in which the left-handed Majorana neutrinos acquire their 
masses via the type I seesaw mechanism. The mass matrices of all Dirac fermions have a similar 
form with four texture zeroes of class I and a normal hierarchy of masses. Then, the mass matrix 
of the left-handed Majorana neutrinos also has a four texture zeroes class I and a normal 
hierarchy of masses. In this scheme, we have a parametrization of the $CKM(PMNS)$ mixing matrix 
in terms of four quark (lepton) mass ratios and one (two) $CP$ violating phase in very good 
agreement with all available experimental data.  Also, with help  of the symmetry group $S_ {3}$ 
we made a classification of mass matrices with texture zeroes in equivalence classes.  In this 
classification we use the similarity transformation $M' = T M T^{-1}$ or $M' = T^{-1} M T$, in 
which we take the transformation matrices $T$  as six elements of  real representation of   
$S_ {3}$. With this classification we reduce the number of phenomenologically viable textures 
for  non-singulars mass matrices of $3\times3$, from thirty 
three down to only eleven independent sets of  matrices . Each of these sets of  matrices has 
exactly the same physical content.
\ack{ We thank Dr. J. Barranco for many inspiring discussions on this problem. 
This work was partially supported by CONACyT Mexico under Contract No. 82291, and DGAPA-UNAM Contract 
No. PAPIIT IN112709.}
\section*{References}
\providecommand{\newblock}{}


\end{document}